\documentclass[12pt]{article}

\usepackage{amsmath,amsthm,amssymb,ascmac}
\usepackage{bm}
\usepackage[english]{babel}

\usepackage{cite}

\newcommand{\g}{{g \hspace{-1.6mm}^{^\circ}}}
\newcommand{\gu}{{{g^{ab}} \hspace{-4.5mm}^{^\circ} \ \ }}
\newcommand{\be}{\beta}
\newcommand{\ka}{\kappa}
\newcommand{\bea}{\begin{eqnarray}}
\newcommand{\eea}{\end{eqnarray}}
\newcommand{\aeq}{\!\!\! &=& \!\!\!}
\newcommand{\aeqd}{\!\!\! &:= &\!\!\!}

\newcommand{\mL}{\mathcal{L}}

\newcommand{\mG}{\mathcal{G}}
\newcommand{\mF}{\mathcal{F}}
\newcommand{\G}{{\rm{G}}}

\newcommand{\defe}{:=}
\newcommand{\e}{\equiv}
\newcommand{\ep}{\varepsilon}

\newcommand{\al}{\alpha}
\newcommand{\bu}{\bullet}

\newcommand{\f}{\frac}
\newcommand{\half}{\frac{1}{2}}
\newcommand{\pr}{\prime}
\newcommand{\dl}{\delta}
\newcommand{\Dl}{\Delta}

\newcommand{\om}{\omega}
\newcommand{\Om}{\Omega}
\newcommand{\com}{\hspace{0.5mm}, \quad}
\newcommand{\la}{\label}
\newcommand{\no}{\nonumber}
\newcommand{\re}[1]{(\ref{#1})}
\newcommand{\res}[1]{\S \ref{#1}}

\newcommand{\RM}[1]{{\rm{#1}}}
\newcommand{\p}{\partial}
\newcommand{\w}{\wedge}
\newcommand{\co}[1]{``{#1}''}

\makeatletter
\@addtoreset{equation}{section}
\makeatother

\title{Generators of local gauge transformations in the covariant canonical formalism of fields}

\author{Satoshi Nakajima\thanks{Department of Physics Engineering, Faculty of Engineering, Mie University, 
Tsu, Mie 514-8507, Japan}
\footnote{subarusatosi@gmail.com}
}

\date{}

\begin{document}

\maketitle

\begin{abstract}

We investigate generators of local gauge transformations in the covariant canonical formalism (CCF) for matter fields, gauge fields and the second order formalism of gravity. 
The CCF treats space and time on an equal footing regarding the differential forms as the basic variables.
The conjugate forms $\pi_A$ are defined as derivatives of the Lagrangian $d$-form $L(\psi^A, d\psi^A)$ with respect to $d\psi^A$, namely $\pi_A := \partial L/\partial d\psi^A$, 
where $\psi^A $ are $p$-form dynamical fields.
The form-canonical equations are derived from the form-Legendre transformation of the Lagrangian form $H:=d\psi^A \wedge \pi_A - L$.
We show that the generator of the local gauge transformation in the CCF is given by 
$\varepsilon^r G_r + d\varepsilon^r \wedge F_r$ where $\varepsilon^r$ are infinitesimal parameters and $G_r$ are the Noether currents which are $(d-1)$-forms.
$\{G_r , G_s \} = f^t_{\ rs}G_t$ holds where $\{\bullet, \bullet \}$ is the Poisson bracket of the CCF and $f^t_{\ rs}$ are the structure constants of the gauge group.
For the gauge fields and the gravity, $G_r=-\{F_r, H \}$ holds. 
For the matter fields, $F_r=0$ holds. 

\end{abstract}

\section{Introduction}
In the traditional analytical mechanics of fields, the canonical formalism gives especial weight to time.
The covariant canonical formalism (CCF) \cite{10, 11, 12, 13, N04, K, 2015, 2016, 2017, 2019, 2020} is a covariant extension of the traditional canonical formalism.
The form-Legendre transformation and the form-canonical equations are derived from a Lagrangian $d$-form with $p$-form dynamical fields $\psi^A$. 
The conjugate forms are defined as derivatives of the Lagrangian form with respect to $d\psi^A$. 
One can obtain the form-canonical equations of gauge theories or those of the second order formalism of gravity without fixing a gauge 
nor introducing Dirac bracket nor any other artificial tricks.
Although the second order formalism of gravity (of which the dynamical variable is only the frame form (vielbein)) 
is a non-constrained system in the CCF, the first order formalism (of which the dynamical variables are both the frame form and the connection form) 
is a constrained system even in the CCF. 
In Refs.\cite{10,11,12,13}, the CCFs of the first order formalism of gravity and supergravity have been studied. 
Only for $d=4$, the CCF of the second order formalism of gravity without Dirac field \cite{K} and with Dirac field \cite{2015} have been studied
\footnote{In Refs.\cite{K, 2015}, the method to derive the form-canonical equations used special characteristics of $d=4$.
Without using those, we derive the form-canonical equations in this paper.
}
.

Poisson brackets of the CCF are proposed in Refs.\cite{10, 2019} and in Ref.\cite{2017} independently.
These are equivalent. 
Although the form-canonical equations of the CCF are equivalent to modified De Donder-Weyl equations \cite{2016}, 
the Poisson bracket of the CCF is not equivalent to it of the De Donder-Weyl theory proposed in Ref.\cite{IV}. 
Reference \cite{2019} introduced the generator of the CCF and studied it of the local Lorentz transformation of gravity in the first order formalism. 
The generators of the local Lorentz transformation and the supersymmetry for supergravity have been studied \cite{2020} in the first order formalism.

The structure of the paper is as follows.
First, we review the covariant canonical formalism (\res{s_rev_CCF}).
Next, we investigate generators of local gauge transformations in CCF for matter fields, gauge fields and the second order formalism of gravity (\res{s_gauge_generator}). 
The total generator is given by $G=\ep^r G_r + d\ep^r \w F_r$ where $\ep^r$ are infinitesimal parameters and $G_r$ are the Noether currents.
The Noether currents satisfy 
$\{G_r , G_s \} = f^t_{\ rs}G_t$ where $\{\bu, \bu\}$ is the Poisson bracket of the CCF and $f^t_{\ rs}$ are the structure constants of the gauge group.
$F_r=0$ holds for the matter fields.   
For the gauge fields and the gravity, $G_r=-\{F_r, H \}$ holds. 
Here, $H$ is the form-Legendre transformation of the Lagrangian form. 
In Appendix \ref{s_Noether}, we review the Noether currents.
In Appendix \ref{s_gauge_rev}, we review the CCF of gauge fields. 
In Appendix \ref{s_Gravity_D}, we apply the CCF to the second order formalism of gravity with Dirac fields for the arbitrary dimension ($d \ge 3 $). 
In Appendix \ref{formula}, several formulas are listed.  

\section{Covariant canonical formalism} \la{s_rev_CCF}

In this section, we review the covariant canonical formalism.

Let us consider $d$ dimension space-time. 
Suppose a $p$-form $\be$ is described by forms $\{\al^i \}_{i=1}^k$. 
If there exists the form $\om_i$ such that $\be$ behaves under variations $\dl \al^i$ as 
$\dl \be = \dl \al^i \w \om_i $, we call $\om_i$ the {\it derivative} of $\be$ by $\al^i$ 
\footnote{In this case, $\be$ is {\it differentiable} by $\al^i$.}
 and denote 
\bea
\f{\p \be}{\p \al^i} \defe  \om_i.
\eea

The Lagrangian $d$-form $L$ is given by $L=\mL \eta$ where $\mL$ is the Lagrangian density and $\eta= \ast 1$ is the volume form 
\footnote{The Hodge operator $\ast$ maps an arbitrary $p$-form $\om=\om_{\mu_1 \cdots \mu_p}dx^{\mu_1}\w \cdots \w dx^{\mu_p}$ $(p=0,1,\cdots,d)$ to a $(d-p)$-form as
\bea
\ast \om = \f{1}{(d-p)!}E_{\ \ \ \ \ \ \ \nu_1 \cdots \nu_{d-p}}^{\mu_1 \cdots \mu_p}\om_{\mu_1 \cdots \mu_p}
dx^{\nu_1} \w \cdots \w dx^{\nu_{d-p}}. \no
\eea
Here, $E_{\mu_1 \cdots \mu_d}$ is the complete anti-symmetric tensor such that $E_{01\cdots d-1}=\sqrt{-g}$ ($g$ is the determinant of the metric $g_{\mu\nu}$). 
And $\ast \ast \om=-(-1)^{p(d-p)}\om$ holds.}
and described by $\psi$ and $d\psi$, $L=L(\psi,d\psi)$, where $\psi$ is a set the forms of the dynamical fields.
For simplicity, we treat $\psi$ as single $p$-form in this section. 
The Euler-Lagrange equation is given by
\bea
\f{\p L}{\p \psi}-(-1)^p d\f{\p L}{\p d\psi} = 0 .\la{EL}
\eea
The above Euler-Lagrange equation has been used since the 1970's \cite{72, 78, 95}.

The {\it conjugate form} $\pi$ is defined by
\bea
\pi \defe \f{\p L}{\p d\psi}.
\eea
This is a $q$-form where $q\defe d-p-1$. 
The {\it Hamilton $d$-form} is defined by
\bea
H(\psi,\pi)\defe d\psi \w \pi-L 
\eea
and described by $\psi$ and $\pi$. 
The variation of $H$ is given by
\bea
\dl H = (-1)^{(p+1)q}\dl \pi \w d\psi -\dl \psi \w\f{\p L}{\p \psi}.
\eea
Then, we obtain
\bea
\f{\p H}{\p \psi}=  -\f{\p L}{\p \psi}\com
\f{\p H}{\p \pi}= (-1)^{(p+1)q}d\psi .
\eea
By substituting the Euler-Lagrange equation \re{EL}, we obtain the {\it canonical equations}
\bea
d\psi  = (-1)^{(p+1)q}\f{\p H}{\p \pi} \com d\pi = -(-1)^p\f{\p H}{\p \psi}.
\eea

The {\it Poisson bracket} proposed in Ref.\cite{2017} is given by
\bea
\{F,G \} = (-1)^{p(f+d+1)} \f{\p F}{\p \psi}\w \f{\p G}{\p \pi}-(-1)^{(d+p-1)(f+1)}\f{\p F}{\p \pi}\w \f{\p G}{\p \psi}. \la{PB}
\eea
Here, $F$ and $G$ are differentiable by $\psi$ and $\pi$, and $F$ is a $f$-form. 
The Poisson bracket proposed in Ref.\cite{2019}, denoted by $\{F,G \}_\RM{F}$, is given by $\{F,G \}_\RM{F}=-\{G,F \}$.
If $F$, $G$ and $H$ are $f$-form, $g$-form and $h$-form respectively and differentiable by $\psi$ and $\pi$, 
\bea
\{G,F \} \aeq -(-1)^{(f+d+1)(g+d+1)}\{F,G \} ,\\
\{F,G \w H \} \aeq \{F,G \} \w H + (-1)^{(f+d+1)g}G \w \{F,H \},
\eea
and
\bea
(-1)^{(f+d+1)(h+d+1)}\{F,\{G,H\}\} \!\!\! &+& \!\!\! (-1)^{(g+d+1)(f+d+1)}\{G,\{H,F\}\}\no\\
\!\!\! &+& \!\!\! (-1)^{(h+d+1)(g+d+1)}\{H,\{F,G\}\}= 0
\eea
hold.
The  canonical equations can be written as
\bea
d\psi  =-\{H, \psi\} \com d\pi=-\{H, \pi\} .\la{CE_PB}
\eea
The fundamental brackets are
\bea
\{\psi,\pi\}=(-1)^{pd} \com \{\pi,\psi\}=-1 \com \{\psi,\psi\}=0=\{\pi,\pi\}.
\eea
If a form $F$ is differentiable by $\psi$ and $\pi$, and $F$ does not depend positively on space-time points, 
\bea
dF \aeq d\psi\w \f{\p F}{\p \psi}+d\pi\w \f{\p F}{\p \pi} \no\\
\aeq (-1)^{(p+1)q}\f{\p H}{\p \pi}\w  \f{\p F}{\p \psi}-(-1)^p\f{\p H}{\p \psi} \w \f{\p F}{\p \pi} \no\\
\aeq -\{H, F\}
\eea
holds.

\section{Generators of local gauge transformations} \la{s_gauge_generator}

Let us consider that an infinitesimal transformation of dynamical fields $\psi^A$ and its conjugate forms $\pi_A$:
\bea
\psi^A \to \psi^A + \dl \psi^A \com \pi_A \to \pi_A + \dl \pi_A. 
\eea
Here, $A$ is the label of the fields.
If there exists $(d-1)$-form $G$ such that
\bea
\dl \psi^A = \{\psi^A, G \} \com \dl \pi_A = \{\pi_A, G \},
\eea
we call $G$ the {\it generator} of the transformation \cite{2019}. 
If a form $F$ is differentiable by $\psi^A$ and $\pi_A$, the transformation of $F$ is given by 
\bea
\dl F = \{F, G\} .
\eea

In this section, we find the generators of the gauge transformations for matter fields (\res{Matter_fields}), 
gauge fields (\res{Gauge_fields}) and the gravitational field within the second order formalism (\res{Gravitational_field}).

\subsection{Matter fields} \la{Matter_fields}

Let us consider that an infinitesimal global gauge transformation of matter fields: 
\bea
\dl \psi^A = \ep^r (\bm{G}_r)^A_{\ B}\psi^B \com \dl L_0 = 0. \la{trans_matter}
\eea
Here, $\ep^r$ are infinitesimal parameters, $\bm{G}_r$ are representations of the generators of a linear Lie group $\mG$ 
and $L_0(\psi^A,d\psi^A)$ is the Lagrangian form of the matter fields $\psi^A$. 
The matrices $\bm{G}_r$ satisfy 
\bea
[\bm{G}_r, \bm{G}_s] = f^t_{\ rs}\bm{G}_t,
\eea
where $[A,B] \defe AB-BA$ and $f^t_{\ rs}$ are the structure constants of $\mG$.
Under the transformation \re{trans_matter}, the conjugate forms $\pi_A$ behave as
\bea
\dl \pi_A = - \ep^r(\bm{G}_r)^B_{\ A}\pi_B . \la{trans_matter_pi}
\eea
The Noether currents \re{N_current} are given by
\bea
G_r^{(0)} \defe (\bm{G}_r)^A_{\ B}\psi^B \w \pi_A. 
\eea
The Noether currents $G_r^{(0)}$ satisfy 
\bea
\{ \psi^A, G_r^{(0)} \} \aeq (\bm{G}_r)^A_{\ B}\psi^B ,\\
\{\pi_A, G_r^{(0)} \} 
\aeq -(\bm{G}_r)^B_{\ A}\pi_B, \\
\{G_r^{(0)}, G_s^{(0)} \}\aeq f^t_{\ rs} G_t^{(0)}.
\eea
The generator of the transformation \re{trans_matter} is given by $\ep^r G_r^{(0)}$.

To generalize \re{trans_matter} to the local gauge transformation, 
$L_0(\psi^A,d\psi^A)$ should be replaced by $L_0(\psi^A,(D\psi)^A)$ where $(D\psi)^A \defe d\psi^A+A^r (\bm{G}_r)^{A}_{\ B}\w \psi^B$ and 
$A^r$ are the gauge fields. 
The forms $\psi^A$ and $\pi_A$ are independent from $A^r$ and $\pi_r$ where $\pi_r$ is the conjugate forms of $A^r$.

\subsection{Gauge fields} \la{Gauge_fields}

Let us consider that the infinitesimal local gauge transformation of the gauge fields:  
\bea
\dl A^r = \ep^s f^r_{\ st}A^t - d\ep^r \com \dl L_1=0. \la{trans_gauge}
\eea
Here, $L_1$ is the Lagrangian form of the gauge fields. 
Under the transformation \re{trans_gauge}, $\pi_r$ behave as
\bea
\dl \pi_r = - \ep^s f^t_{\ sr} \pi_t .
\eea
The Noether currents \re{N_current} are given by
\bea
G_s^{(1)} \defe f^r_{\ st}A^t \w \pi_r .
\eea
The Noether currents $G_r^{(1)}$ satisfy 
\bea
\{ A^s, G_r^{(1)}\} \aeq  f^s_{\ rt}A^t ,\\
\{\pi_s, G_r^{(1)} \}\aeq - f^t_{\ rs} \pi_t, \\
\{G_r^{(1)}, G_s^{(1)} \} \aeq f^t_{\ rs} G_t^{(1)}.
\eea
We put $G_r \defe G_r^{(0)} +G_r^{(1)}$. 
The Noether currents $G_r$ satisfy 
\bea
\{G_r , G_s \} = f^t_{\ rs}G_t.
\eea
The generator of the transformation without $d\ep^r$ is given by $ \ep^r G_r $.  
We assume that the generator of the local gauge transformation (denoted by $G$) is given by 
\bea
G = \ep^r G_r + d\ep^r \w F_r,
\eea
where $F_r$ are unknown $(d-2)$-forms described by only $\pi_r$. 
Because 
\bea
\{ A^s, d\ep^r \w F_r \} = d\ep^r \w \f{\p F_r}{\p \pi_s}
\eea
holds, we obtain
\bea
F_r = -\pi_r.
\eea
Then, $G$ is given by
\bea
G = \ep^r G_r - d\ep^r \w \pi_r.
\eea
$F_r$ does not affect to $\psi^A$, $\pi_A$ and $\pi_r$:
\bea
\dl \psi^A \aeq \{\psi^A ,G \} = \ep^r (\bm{G}_r)^A_{\ B}\psi^B ,\\
\dl A^r \aeq \{A^r, G \} = \ep^s f^r_{\ st}A^t - d\ep^r ,\\
\dl \pi_A \aeq \{\pi_A ,G \} =-\ep^r(\bm{G}_r)^B_{\ A}\pi_B ,\\
\dl \pi_r \aeq \{\pi_r, G \} =-\ep^s f^t_{\ sr} \pi_t.
\eea

\subsection{Gravitational field} \la{Gravitational_field}

\subsubsection{Notation}

We explain the notations used in this paper.
Let $g$ be the metric of which has signature $(-+\cdots+)$, and let $\{\theta^a\}_{a=0}^{d-1}$ denote an orthonormal frame (vielbein).
We have $g=\g_{ab} \theta^a \otimes \theta^b$ with $\g_{ab}\defe \RM{diag}(-1,1,\cdots,1)$.
All indices are lowered and raised with $\g_{ab}$ or its inverse $\gu$.
The first structure equation is 
\bea
d\theta^a+\om^a_{\ b}\w \theta^b=\Theta^a , \la{FS}
\eea
where $\om^a_{\ b}$ is the connection form and  $\Theta^a$ is the torsion 2-form. 
In the following of this paper, we suppose $\om_{ba}=-\om_{ab}$. 
We put
\bea
\eta^a =\ast \theta^a,\ \eta^{ab}=\ast(\theta^a \w \theta^b), \ \eta^{abc}=\ast(\theta^a \w \theta^b \w \theta^c), \ \eta^{abcd}=\ast(\theta^a \w \theta^b \w \theta^c \w \theta^d) .
\eea
In Appendix \ref{formula}, several identities about 
$\theta^a \w \eta_{a_1 \cdots a_r}$ $(r=1,2,3,4)$ , $\dl \eta_{a_1 \cdots a_r}(r=0,1,2,3)$ and $d\eta_{a_1 \cdots a_r}$ $(r=1,2,3)$  are listed.

\subsubsection{Generators of local Lorentz transformation}

Let us consider that an infinitesimal local Lorentz transformation
\bea
\dl \theta^a = \ep^a_{\ b}\theta^b .\la{trans_gravity}
\eea
Here, $\ep^{ab}$ are infinitesimal parameters which satisfy $\ep^{ab}=-\ep^{ba}$. 
Under the transformation, $\om^{ab}$ behave as
\bea
\dl \om^{ab} = \ep^a_{\ c}\om^{cb}+\ep^{b}_{\ c}\om^{ac}-d\ep^{ab}. 
\eea
Using \re{C10}, the conjugate form of $\theta^a$ is given by
\bea
\pi_a=\f{1}{2\ka}\om^{bc} \w \eta_{abc}. \la{def_pi} 
\eea
Here, $\ka$ is the Einstein constant. 
$\pi_a$ behave as 
\bea
\dl \pi_a \aeq \f{1}{2\ka}\dl \om^{bc}\w \eta_{abc}+\f{1}{2\ka} \om^{bc}\w \dl \theta^d \w \eta_{abcd} \no\\
\aeq \ep^{bc}(- \g_{a[c} \pi_{b]}) -d\ep^{bc} \w \f{1}{2\ka} \eta_{abc}.
\eea
Here, we used \re{A11} in the first line and \re{A0} in the second line. 
The Noether currents \re{N_current} are given by
\bea
G_{cd} \defe 2\theta_{[d}\w \pi_{c]}=\theta_{d}\w \pi_{c}-\theta_{c}\w \pi_{d}.
\eea
The Noether currents $G_{cd}$ satisfy 
\bea
\{\theta^a, G_{cd} \} \aeq  2\theta_{[d} \dl^a_{c]} ,\\
\{ \pi_a, G_{cd} \} \aeq -2\g_{a[d} \pi_{c]}, \\
\{G_{ab}, G_{cd} \}\aeq \g_{bc}G_{ad}-\g_{ac}G_{bd}+\g_{ad}G_{bc}-\g_{bd}G_{ac} \la{G_LG}. 
\eea
\re{G_LG} corresponds to the commutation relations of the generators of the Lorentz group:
\bea
[\bm{G}_{ab}, \bm{G}_{cd}] = \g_{bc}\bm{G}_{ad}-\g_{ac}\bm{G}_{bd}+\g_{ad}\bm{G}_{bc}-\g_{bd}\bm{G}_{ac}.  
\eea
The generator of the transformation without $d\ep^{ab}$ is given by $\half \ep^{ab}G_{ab}$. 
We assume that the generator of the local Lorentz transformation \re{trans_gravity} (denoted by $G$) is given by 
\bea
G = \half \ep^{ab}G_{ab} + \half d\ep^{ab} \w F_{ab},
\eea
where $F_{ab}$ are $(d-2)$-forms described by only $\theta^a$. 
Because 
\bea
\{\pi_a, \half d\ep^{bc} \w F_{bc}  \} = \half d\ep^{bc}\w \f{\p F_{bc}}{\p \theta^a}
\eea
holds and the right hand side of the above equation should be $-d\ep^{bc} \w \f{1}{2\ka} \eta_{abc}$, we obtain
\bea
F_{bc} = -\f{1}{\ka}\eta_{bc}.
\eea
We used \re{A5}. 
Then, $G$ is given by
\bea
G = \half \ep^{ab}G_{ab} - \f{1}{2\ka} d\ep^{ab} \w \eta_{ab}.
\eea
$F_{ab}$ does not affect to $\theta^a$:
\bea
\dl \theta^a \aeq \{\theta^a, G \} = \ep^a_{\ b}\theta^b ,\\
\dl \pi_a \aeq \{\pi_a, G \}= -\ep^{b}_{\ a}\pi_{b}-d\ep^{bc} \w \f{1}{2\ka} \eta_{abc}.
\eea

\subsection{The relation between $F_r$ and $G_r$}

According to Ref.\cite{2019}, if a generator is given by $G =\ep^r G_r +d\ep^r \w F_r$ with nonzero $F_r$,  
\bea
G_r = -\{F_r, H \}
\eea
holds. 
We check this relationship for the gauge fields and the gravitational field.
For the gauge fields, 
\bea
-\{F_r, H_1 \} \aeq \{\pi_r, H_1 \} \no\\
\aeq -\f{\p H_1}{\p A^r} \no\\
\aeq f^c_{\ rb}A^b \w \pi_c \no\\
\aeq G_r^{(1)}
\eea
holds. 
Here, $H_1$ is the Hamilton form of the gauge fields \re{H_1} and we used \re{p_H_1} in the third line.
For the gravitational field, 
\bea
-\{F_{ab}, H_\G \} \aeq \f{1}{\ka}\{\eta_{ab}, H_\G \} \no\\
\aeq -\f{1}{\ka} \f{\p \eta_{ab}}{\p \theta^c} \w \f{\p H_\G}{\p \pi_c} \no\\
\aeq -\f{1}{\ka}  \eta_{abc} \w (-\om^c_{\ d}\w \theta^d) \no\\
\aeq \f{1}{\ka} \om^c_{\ d} \w \theta^d \w \eta_{abc} 
\eea
holds. 
In the third line, we used \re{p_H_G} and \re{A5}. 
Because 
\bea
\ka G_{ab} \aeq - \om^{cd} \w \theta_{[b} \w \eta_{a]cd} =-2\om^c_{\ [b} \w \eta_{a]c} ,\\
-\ka \{F_{ab}, H_\G \} \aeq \om^c_{\ d} \w  \theta^d \w \eta_{abc} =-2\om^c_{\ [b}\w \eta_{a]c}
\eea
hold using \re{A1}, we have
\bea
G_{ab} = -\{F_{ab}, H_\G \}. 
\eea

\section*{Acknowledgment} 
We acknowledge helpful discussions with M. Matsuo.

\appendix

\section{Noether currents} \la{s_Noether}

We explain the Noether currents.
For an infinitesimal transformation of $p$-form dynamical fields $\psi^A \to \psi^A + \dl \psi^A$, an identical equation
\bea
\dl L \e \dl \psi^A \w \Big( \f{\p L}{\p \psi^A}-(-1)^p d\f{\p L}{\p d\psi^A} \Big)
+d\Big(\dl \psi^A \w \f{\p L}{\p d\psi^A} \Big) \la{delta_L_2}
\eea
holds. 
Here, $\e$ denotes identical equation which holds without using the Euler-Lagrange equations. 
For a global transformation
\bea
\dl \psi^A = \ep^r \Dl^A_r \com
\dl L= \ep^r dl_r,
\eea
we have
\bea
d l_r \e \Dl^A_r \w [L]_A+d\Big(\Dl^A_r \w \f{\p L}{\p d\psi^A} \Big) 
\eea
from \re{delta_L_2}. 
Here, $[L]_A \defe \p L/\p \psi^A-(-1)^p d(\p L/\p d\psi^A)$. 
Under the Euler-Lagrange equations $[L]_A=0$, the Noether currents
\bea
N_r \defe \Dl^A_r \w \f{\p L}{\p d\psi^A} -l_r \la{N_current}
\eea
are conserved: $dN_r= 0$.

\section{Covariant canonical formalism of gauge fields} \la{s_gauge_rev}

In this section, we review that the covariant canonical formalism of gauge fields.

The curvatures of the gauge fields are defined by 
\bea
\mF^r \defe dA^r +\half f^r_{\ bc}A^b \w A^c.
\eea
We put $\mF_r \defe \ka_{rs}\mF^s$ with the Killing form $\ka_{rs}\defe -f^a_{\ rb}f^b_{\ sa}(=\ka_{sr})$. 
The Lagrangian form of the gauge fields $L_1$ is given by 
\bea
L_1 = -\f{1}{2k} \mF^r \w \ast \mF_r, \la{L_gauge}
\eea
where $k$ is a positive constant.
$L_1$ is gauge invariant.

We consider the Euler-Lagrange equation of the gauge fields. 
The derivatives of the total Lagrangian form 
$L \defe L_0(\psi^A,(D\psi)^A)+L_1$ are given by
\bea
\f{\p L}{\p A^a} = -\f{1}{k}f^c_{\ ab}A^b \w \ast \mF_c+J_a \com  \f{\p L}{\p dA^a} = -\f{1}{k}\ast \mF_a
\eea
with
\bea
J_a \defe \f{\p L_0(\psi^A,(D\psi)^A)}{\p A^a}.
\eea
The Euler-Lagrange equation $\p L/\p A^a+d(\p L/\p dA^a)=0$ is given by
\bea
D\ast \mF_a\defe d\ast \mF_a+f^c_{\ ab}A^b \w \ast \mF_c =kJ_a \la{YMU_CAM}.
\eea
$D\ast \mF_a$ is the covariant derivative. 
The above equation is the Yang-Mills-Utiyama equation.

The conjugate form of $A^a$ is given by
\bea
\pi_a=-\f{1}{k}\ast \mF_a.
\eea
The Hamilton form is given by
\bea
H\aeq H_1 -L_0(\psi^A,(D\psi)^A) ,\\
H_1 \aeqd -\f{1}{2} f^a_{\ bc} A^b \w A^c\w \pi_a+\f{k}{2} \pi_a \w \ast \pi^a, \la{H_1} 
\eea
with $\pi^a \defe (\ka^{-1})^{ab}\pi_b= -\f{1}{k}\ast \mF^a$. 
The derivatives of $H$ are given by
\bea
\f{\p H}{\p A^a} \aeq -f^c_{\ ab}  A^b \w \pi_c-J_a ,\la{p_H_1}\\ 
\f{\p H}{\p \pi_a} \aeq k \ast \pi^a-\f{1}{2} f^a_{\ bc}A^b \w A^c. 
\eea
The canonical equations $dA^a=\f{\p H}{\p \pi_a}$ and $d\pi_a = \f{\p H}{\p A^a}$ become 
\bea
dA^a \aeq k \ast \pi^a-\f{1}{2} f^a_{\ bc}A^b \w A^c,\\
d\pi_a \aeq -f^c_{\ ab}  A^b \w \pi_c-J_a .
\eea
The above two equations can be rewritten as 
\bea
\mF^a = k\ast \pi^a \com D\pi_a \defe d\pi_a+f^c_{\ ab}  A^b \w \pi_c=-J_a.
\eea
The former is equivalent with the definition of $\pi_a$.
The latter is  equivalent with the Yang-Mills-Utiyama equation.
The covariant canonical formalism does not need the gauge fixing.

\section{Second order formalism of gravity for $d \ge 3$} \la{s_Gravity_D}

In this section, we apply the CCF to the second order formalism of gravity with Dirac fields for the arbitrary dimension ($d \ge 3)$.

\subsection{Notation}

The curvature 2-form $\Om^a_{\ b}$ is given by 
\bea
\Om^a_{\ b}=d\om^a_{\ b}+\om^a_{\ c}\w \om^c_{\ b} .
\eea
Expanding the curvature form as
\bea
\Om^a_{\ b}=\half R^a_{\ bcd}\theta^c \w \theta^d ,
\eea
we define 
\bea
R_{ab}\defe R^c_{\ acb}, \ R\defe R^a_{\ a} .
\eea
Because of \re{A4},  $R \eta$ can be written as
\bea
 R \eta = \Om^{ab} \w \eta_{ab} \la{ast_R}.
\eea

\subsection{Lagrange formalism} \la{L_F}

The Lagrangian form of the gravity in the second order formalism is given by
\bea
L(\theta,d\theta) = L_\G(\theta,d\theta)+L_\RM{mat}(\theta,d\theta).
\eea
Here, $L_\G$ is the Lagrangian form for the pure gravity given by
\bea
L_\G(\theta,d\theta) = \f{1}{2\ka}N^\pr \com N^\pr \defe R \eta-d(\om^{ab} \w \eta_{ab}) \la{def_N^pr},
\eea
and $L_\RM{mat}(\theta,d\theta)=L_\RM{mat}(\theta,\om(\theta,d\theta))$ is the Lagrangian form of \co{matters} which are scalar fields, Dirac fields and gauge fields.
Here, $\ka$ is the Einstein constant. 
Only the Dirac fields couple to $\om^{ab}$.

We derive the Euler-Lagrange equation of the gravity.
The variation of $L$ is given by
\footnote{$\dl N^\pr= \dl(\om^a_{\ c} \w \om^{cb} \w \eta_{ab} +\om^{ab} \w d \eta_{ab})$ and 
\bea
\om^{ab} \w  \dl d \eta_{ab} = \dl d\theta^c \w \om^{ab} \w \eta_{abc}+ \dl \theta^c \w \om^{ab} \w d\eta_{abc}  \no
\eea
hold. 
Here, we used the following formulas:
\bea
d\eta_{ab}=d\theta^c \w \eta_{abc} \com \dl \eta_{abc}=\dl \theta^d \w \eta_{abcd} \com d\eta_{abc}=d\theta^d \w \eta_{abcd}. \no
\eea
}
\bea
\dl L(\theta,d\theta) \aeq \dl \theta^c \w \Big( \f{1}{2\ka}[\Om^{ab} \w \eta_{abc} -d(\om^{ab} \w \eta_{abc})]+T_c \Big)
+ \dl d \theta^c \w \f{1}{2\ka} \om^{ab} \w \eta_{abc} \no\\
&&+ \dl \om^{ab}(\theta,d\theta) \w \Big(\f{1}{2\ka}[d\eta_{ab}-\om^c_{\ a} \w \eta_{cb} -\om^c_{\ b}\w \eta_{ac}]+\f{\p L_\RM{mat}}{\p \om^{ab}} \Big), \la{dl_L}
\eea
where 
\bea
T_a \defe \f{\p L_\RM{mat}(\theta,\om)}{\p \theta^a}
\eea
is the energy-momentum form. 
We suppose that
\bea
\f{1}{2\ka}[d\eta_{ab}-\om^c_{\ a} \w \eta_{cb} -\om^c_{\ b}\w \eta_{ac}]+\f{\p L_\RM{mat}}{\p \om^{ab}} \e 0 ,\la{de_ab2}
\eea
which is the same as the Euler-Lagrange equation for the connection of the first order formalism 
\footnote{
In the second order  formalism, \re{de_ab2} is not an Euler-Lagrange equation. 
$\om^{ab}$ is determined by \re{de_ab2}. 
The method to obtain \re{C10} in this subsection is also called the 1.5 order  formalism.
}
.
Under this supposition, \re{dl_L} leads to
\bea
\f{\p L}{\p \theta^c}=  \f{1}{2\ka}[\Om^{ab} \w\eta_{abc} -d(\om^{ab} \w \eta_{abc} )]+T_c \com 
\f{\p L}{\p d\theta^c}= \f{1}{2\ka} \om^{ab}\w \eta_{abc}. \la{C10}
\eea
The Euler-Lagrange equation $\p L/\p \theta^c+d(\p L/\p d\theta^c)=0$ becomes the Einstein equation
\bea
-\f{1}{2\ka}\Om^{ab} \w\eta_{abc} = T_c . \la{Einstein_equation}
\eea
If we expand $T_c$ as $T_c=T_c^{\ b} \eta_b$, the above  equation leads to
\bea
R^a_{\ b} -\half R\dl^a_b = \ka T_b^{\ a}.
\eea

\subsection{Covariant canonical formalism} \la{H_F}

Next, we consider the covariant canonical formalism.

In \re{def_N^pr}, $N^\pr$ can be rewritten as 
\bea
N^\pr \aeq \om^a_{\ c}\w \om^{cb}\w \eta_{ab}+ \om^{ab}\w d\eta_{ab} \la{def_N2} \no\\
\aeq N+\Theta^a \w \om^{bc} \w \eta_{abc} \la{N^pr_N}
\eea
with 
\bea
N \defe \om^a_{\ c}\w \om^{cb} \w \eta_{ba}. \la{def_N_general}
\eea
In the second line of \re{N^pr_N}, we used \re{A9}.
Using \re{FS} and \re{A1}, $N$ can be rewritten as 
\bea
N = d\theta^a \w \half \om^{bc} \w \eta_{abc}-\Theta^a \w \half \om^{bc} \w \eta_{abc}. \la{N_pi}
\eea

The conjugate form of $\theta^a$ is given by \re{def_pi}. 
The Hamilton form is given by
\bea
H(\theta,\pi) = d\theta^a \w \pi_a-L=H_\G(\theta,\pi)-L_\RM{mat}(\theta,\pi) 
\eea
with
\bea
H_\G(\theta,\pi) \defe \f{N}{2\ka} .
\eea
Here, we used \re{N^pr_N} and \re{N_pi}.
Because $C_{abc}$ is described by the Dirac fields, it is independent from $\theta^a$ and $\pi_a$. 
Then, $\Theta^a=\half C^a_{\ bc}\theta^b\w \theta^{c}$ is independent from $\pi_a$.

The canonical equations are given by 
\bea
d\theta^a \aeq \f{\p H_\G}{\p \pi_a} -\f{\p L_\RM{mat}}{\p \pi_a} ,\la{C_eq_theta}\\
d\pi_a \aeq \f{\p H_\G}{\p \theta^a}-\f{\p L_\RM{mat}(\theta,\pi)}{\p \theta^a}. \la{C_eq_Pi}
\eea
In the right hand side of \re{C_eq_theta}, the second term can be rewritten as
\bea
-\f{\p L_\RM{mat}}{\p \pi_a} =  \Theta^a
\eea
using \re{de_ab2}. 
Then, \re{C_eq_theta} becomes 
\bea
d\theta^a = \f{\p H_\G}{\p \pi_a}+\Theta^a \la{H1a}.
\eea

To calculate $\p H_\G/\p \pi_a$, $\p H_\G/\p \theta^a$, 
we represent $N$ by $\theta^a$ and $\pi_a$.
Using \re{A4}, $N$ can be rewritten as
\bea
N = (\om_{abc}\om^{bca}+\om_a \om^a)\eta.
\eea
Here, we expand $\om_{ab}$ as $\om_{ab}=\om_{abc}\theta^c$ and put $\om_a \defe \om^b_{\ ab}$.
We can represent $\om_{abc}$ by $\theta^a$ and $\pi_a$ as
\bea
\om_{abc} \aeq \ka \Big[ v_{c,ab}+\f{1}{d-2}(\g_{ac} v_b - \g_{bc} v_a) \Big] ,\la{om_v}\\
v_{c,ab} \aeqd -\ast V_{c,ab} \com V_{c,ab}\defe \pi_c\w \theta_a \w \theta_b
\eea
with $v_a \defe v^{b}_{\ ab}$.

Next, we derive the canonical equations. 
For an arbitrary $d$-form $\xi$, 
\bea
\dl v_{c,ab} \xi \aeq - \dl v_{c,ab} \xi \ast \eta =- \dl v_{c,ab}\eta \ast \xi = (- \dl[ v_{c,ab}\eta]+ v_{c,ab}\dl \eta )\ast \xi \no\\
\aeq (-\dl V_{c,ab} + v_{c,ab}\dl \eta )\ast \xi
\eea
holds \cite{K}. Then, we have
\bea
\dl v_{c,ab} \eta = \dl V_{c,ab} - v_{c,ab}\dl \eta .\la{tot_va}
\eea
Using this, we have
\bea
\f{\p H_\RM{G}}{\p \pi_c} \aeq -\om^c_{\ a} \w \theta^a , \la{p_H_G} \\
 \f{\p H_\G}{\p \theta^d} \aeq  \f{1}{2\ka}(\om^{c}_{\ d} \w \om^{ab}\w \eta_{abc}+ \om^a_{\ c}\w \om^{bc} \w \eta_{bad}). \la{Goal}
\eea
Substituting \re{p_H_G} into \re{C_eq_theta}, we have
\bea
d\theta^a = -\om^a_{\ b}\w \theta^b + \Theta^a ,\la{G_CE1}
\eea
which is equivalent to the first structure equation \re{FS}.
By the way, using \re{de_ab2}, we can show that
\bea
-\f{\p L_\RM{mat}(\theta,\pi)}{\p \theta^c} = -T_c - \f{1}{2\ka}\om^{ab} \w\Theta^d \w \eta_{abcd} . \la{p_L_mat_pi}
\eea
Substituting  \re{Goal} and \re{p_L_mat_pi} into \re{C_eq_Pi}, we have
\bea
d\pi_c =  \f{1}{2\ka}(\om^d_{\ b} \w \om^{ab}\w \eta_{adc}+ \om^d_{\ c} \w \om^{ab}\w \eta_{abd} -\om^{ab}\w \Theta^d \w \eta_{abcd})-T_c . \la{G_CE2}
\eea

\section{Formulas} \la{formula}

Several useful formulas are listed. 
For $\theta^a \w \eta_{a_1 \cdots a_r}(r=1,2,3,4)$, 
\bea
\theta^b \w \eta_{a_1\cdots a_r} \aeq (-1)^{r-1} r \dl^b_{[a_1}\eta_{a_2 \cdots a_r]} \la{A0}\la{A1}\la{A2}\la{A3}
\eea
hold \cite{95}. 
Using this, we have
\bea
\theta^a \w \theta^b \w \eta_{cd} \aeq (\dl^a_c\dl^b_d-\dl^a_d\dl^b_c)\eta. \la{A4} 
\eea
For $\dl \eta_{a_1 \cdots a_r}(r=0,1,2,3)$, 
\bea
\dl \eta_{a_1\cdots a_r} \aeq \dl \theta^b \w \eta_{a_1\cdots a_r b}  \la{A11} \la{A5} \la{A6} \la{A7}
\eea
hold.
For $d\eta_{a_1 \cdots a_r}(r=1,2,3)$,
\bea
d \eta_{a_1\cdots a_r} 
\aeq (r+1)\om^b_{\ [a_1}\w \eta_{ba_2 \cdots a_r]}+ \Theta^b \w \eta_{a_1\cdots a_r b} \la{A8} \la{A9}\la{A10}
\eea
hold. 
$\eta_{a_1 \cdots a_r}(r=1,2,3,4)$ can be written as \cite{95}
\bea
\eta_{a_1 \cdots a_r} \aeq e_{a_r} \rfloor \eta_{a_1 \cdots a_{r-1}}. \la{A15}\la{A16}\la{A17}\la{A18}
\eea
Here, $\rfloor$ is the interior product and $\{e_a\}$ is the dual basis of $\{\theta^a \}$ ($ e_a \rfloor \theta^b = \dl_a^b$).

\end{document}